\documentclass[a4paper, 12pt]{article}
\usepackage{amsmath}
\numberwithin{equation}{section}
\begin{document}
\begin{flushright}
    hep-th/0602118 \\
    DAMTP-2006-14
  \end{flushright}
\vskip 1cm
  \begin{center}
\LARGE{\textbf{A Killing tensor for higher dimensional Kerr-AdS black
    holes with NUT charge }}
\end{center}
\begin{center}
        \begin{center}
        {\Large Paul Davis\footnote{P.Davis@damtp.cam.ac.uk} }\\
        \bigskip\medskip
        {\it  DAMTP, Centre for Mathematical Sciences,\\
      University of Cambridge,
      Wilberforce Rd.,\\
      Cambridge CB3 0WA, UK\\}
        \end{center}
  \end{center}
  \vskip 0.5cm


\begin{abstract}
In this paper, we study the recently discovered family of higher
dimensional Kerr-AdS black holes with an extra NUT-like parameter. We
show that the inverse metric is additively separable after
multiplication by a simple function. This allows us to separate the
Hamilton-Jacobi equation, showing that geodesic motion is integrable
on this background. The separation of the Hamilton-Jacobi equation is
intimately linked to the existence of an irreducible Killing tensor,
which provides an extra constant of motion. We also demonstrate that
the Klein-Gordon equation for this background is separable.   
\end{abstract} 


\section{Introduction}
The idea of separability within the study of solutions to the
Einstein equations has proved extremely fruitful. The four dimensional
Kerr-(A)dS black hole was discovered by Carter in \cite{carter}, by
considering solutions of Einstein's equations that permit separable
solutions to the Hamilton-Jacobi and Klein-Gordon equations. Such
separable solutions possess a hidden symmetry which is intimately
related to the existence of a non-trivial Killing tensor. This is a
second rank tensor, $K^{\mu \nu}$, that satisfies a generalised
version of Killing's equation, $\nabla_{(\mu} K_{\nu \sigma)} =
0$. Such a Killing tensor gives rise to an additional constant of
motion that is quadratic in the canonical momenta. This additional
constant of motion, together with those arising from the Killing
vectors, renders geodesic motion integrable, and the geodesic equation
solvable by quadratures. Such integrable dynamical systems are rather
rare and it is therefore a valuable exercise to study them in some detail.  

\vskip 0.2cm
\noindent
With the emergence of string theory as a potential theory
of everything, higher dimensional analogues of the four dimensional
black holes were sought. The higher dimensional analogues of the
asymptotically flat Kerr black hole were found by Myers and Perry,
\cite{mp}, and it was only much more recently that higher
dimensional, Kerr-(A)dS black holes were constructed, \cite{glpp}. The
five dimensional Myers-Perry and Kerr-(A)dS black holes were shown to
possess Killing tensors for arbitrary rotation parameters in
\cite{stoj} and \cite{kl1} respectively, and in the special case where
there are two sets of equal rotation parameters, a Killing tensor has
been found in all dimensions, \cite{vs}.  

\vskip 0.2cm
\noindent
These higher dimensional, rotating black holes can be further
generalised to include various types of charge. Electrically
charged, Kerr-AdS black hole solutions have only recently been found
in the context of $D=5$ minimal gauged supergravity. The case where
the two rotation parameters are equal was discovered in \cite{clp4}
and shown to have a reducible Killing tensor in \cite{kl2}. This
Killing tensor is also found in the case of black holes with equal
rotation parameters in the more general $U(1)^3$  theory. The case of
arbitrary rotation parameters was found in \cite{cclp} and was shown
to have an irreducible Killing tensor in \cite{dkl}. 

\vskip 0.2cm
\noindent
Another generalisation, and the one under consideration in
this paper, is to include a NUT-like parameter. NUT charged spacetimes
are of interest for many reasons. For example, their Euclidean
extensions can be used in the study of monopoles in gauge theories and
in the context of the AdS/CFT correspondence, asymptotically AdS, NUT
charged spacetimes can be used to study various issues in the field of
chronology protection. The four-dimensional Kerr-Taub-NUT-de Sitter
solution was found by Carter in \cite{carter2} using the idea of
separability. Higher dimensional versions with one non-zero rotation
parameter were found and shown to possess a Killing tensor in
\cite{cglp}. A large class of Taub-NUT solutions were shown to possess
a Killing tensor in higher dimensions in \cite{v}.  In \cite{clp1}, a
class of cohomogeneity-2, rotating, AdS black holes in higher
dimensions were found that possess a generalisation of the NUT
parameter, and it is these solutions that will provide the focus of
this paper.   

\vskip 0.2cm
\noindent
The metric takes slightly different forms in odd and even dimensions,
but in both cases, the inverse metric multiplied by a simple function
can be separated (non-uniquely) into the sum of two parts, each
depending on a different coordinate. This allows the Hamilton-Jacobi
equation to admit an additively separable solution, establishing geodesic
integrability. This separation introduces a separation constant which
can be understood as arising from a second rank Killing tensor,
$K^{\mu \nu}$, contracted with the canonical momenta. This Killing
tensor is found and shown to have a smooth limit as the cosmological
constant tends to zero. We also show that the Klein-Gordon equation
has separable solutions in all dimensions. 

\vskip 0.2cm
\noindent 
The structure of this paper is as follows. In Section 2, we study the
odd-dimensional Kerr-AdS-NUT Solutions. We present the inverse metric
and show that some multiple of it is additively separable as a
function of the radial variable and an angular variable. We then
demonstrate that the Hamilton-Jacobi equation is separable and extract
the Killing tensor from the separated angular equation. The last part
of Section 2 demonstrates that the Klein-Gordon equation will possess
separable solutions. Section 3 follows the same procedure as Section
2, but we consider the even dimensional case. In Section 4 we briefly
summarise the main results of this work. 

\vskip 0.2cm
\noindent 
As this work was being written up, \cite{clp3} appeared on
the archive and showed exactly the same results as demonstrated here,
namely the separability of the Hamilton-Jacobi equation and the
Klein-Gordon equation, but in a different coordinate system to that
used here.


\section{Odd Dimensional Kerr-AdS-NUT Solutions}
In odd spacetime dimensions, $D= 2n +1$, the general Kerr-AdS black hole
possesses $n$ latitude coordinates $\mu_i$ subject to $\sum_{i=1}^n
\mu_i^2 =1$, $n$ azimuthal angles, a radial coordinate, $r$ and a time
coordinate, $t$. The generalisation that includes a NUT-type parameter
requires a restriction on the $n$ angular momenta so that they form
two sets, with $p$ angular momenta taking the value $a$ and $q$ taking
the value $b$. In order to make this more transparent, a new
coordinate, $\theta$ is introduced such that 
\begin{eqnarray} \label{lat}
\mu_i &=& \nu_i \sin \theta, \quad 1 \leq i \leq p, \quad \sum_{i = 1}^p
\nu_i^2 = 1, \nonumber \\
\mu_{j + p} &=& \tilde{\nu}_j \cos \theta, \quad 1 \leq j \leq q,
\quad \sum_{j = 1}^q \tilde{\nu}_j^2 = 1,
\end{eqnarray}
where $n = p + q$. It is convenient not to use the coordinate,
$\theta$, but to use the related variable, $v$, which satisfies
\begin{equation}
v^2 = a^2 \cos^2 \theta + b^2 \sin^2 \theta.
\end{equation}
For convenience, the azimuthal coordinates are also partitioned into
two sets, with $p$ of them denoted $\phi_i$ and $q$ of them denoted
$\tilde{\phi}_j$. 

\vskip 0.2cm
\noindent
In terms of these new coordinates, the metric for the Kerr-AdS spacetime
with the extra NUT-type parameter can be written \cite{clp1}
\begin{eqnarray}
d s^2 &=& - \frac{ \Delta_r \Delta_v }{\Xi_a \Xi_b} d t^2 +
\frac{\rho^{2n -2} d r^2}{U} + \frac{ \omega^{2n-2} d v^2}{V}
+ \frac{2 M}{\rho^{2n-2}} \Bigg( \frac{ \Delta_v \, d t}{\Xi_a \Xi_b}
 - \mathcal{A} \Bigg)^2 + \frac{2 L}{\omega^{2n-2}} \Bigg( \frac{
 \Delta_r \, d t}{\Xi_a \Xi_b} - \tilde{\mathcal{A}} \Bigg)^2
\nonumber \\
&+& \frac{ (r^2 + a^2)(a^2 - v^2)}{\Xi_a (a^2 - b^2)} \sum_{i = 1}^p
\big( d \nu_i^2 + \nu_i^2 d \phi_i^2 \big) + \frac{ (r^2 + b^2)(b^2 -
  v^2)}{\Xi_b  (b^2 - a^2)} \sum_{j = 1}^q
\big( d \tilde{\nu_j}^2 + \tilde{\nu}_j^2 d \tilde{\phi}_j^2 \big) ,
\end{eqnarray}
where 
\begin{eqnarray}
\mathcal{A} &=& \frac{ a (a^2 - v^2)}{\Xi_a (a^2 - b^2)} \sum_{i = 1}^p
\nu_i^2 d \phi_i + \frac{ b (b^2 - v^2)}{\Xi_b (b^2 - a^2)} \sum_{j = 1}^q
\tilde{\nu}_j^2 d \tilde{\phi}_j, \nonumber \\
\tilde{\mathcal{A}} &=& \frac{ a (r^2 + a^2)}{\Xi_a (a^2 - b^2)}
\sum_{i = 1}^p \nu_i^2 d \phi_i + \frac{ b (r^2 + b^2)}{\Xi_b (b^2 -
  a^2)} \sum_{j = 1}^q \tilde{\nu}_j^2 d \tilde{\phi}_j, \nonumber \\
U(r) &=& \frac{\Delta_r (r^2 + a^2)^p (r^2 + b^2)^q}{r^2} - 2 M,
\nonumber \\
V(v) &=& - \frac{\Delta_v (a^2 - v^2)^p (b^2 - v^2)^q}{v^2}  + 2
L, \nonumber \\
\rho^{2n-2} &=& (r^2 + v^2) (r^2 + a^2)^{p-1} (r^2 + b^2)^{q-1},
\nonumber \\
\omega^{2n-2} &=& (r^2 + v^2) (a^2 - v^2)^{p-1} (b^2 - v^2)^{q-1},
\nonumber \\
\Delta_r &=& 1 + g^2 r^2, \quad \Delta_v = 1 - g^2 v^2,\nonumber \\
\Xi_a &=& 1  - a^2 g^2, \quad \textrm{and} \quad \Xi_b = 1 - g^2 b^2.
\end{eqnarray}

\vskip 0.2cm
\noindent
In order to investigate the separability of the Hamilton-Jacobi and
Klein-Gordon equations in this background, we need to compute the
inverse metric and show that, after being multiplied by a simple
function, the $r$ and $v$ dependence of each component is additively
separable. Performing this computation, we find that the inverse metric
is given by: 
\begin{eqnarray}
(r^2 + v^2) g^{t t} &=& \frac{a^2 b^2 g^2 + r^2 ( g^2 (a^2 + b^2)
    -1)}{\Delta_r } + \frac{b^2 (a^2 - v^2) \Xi_b - a^2 (b^2 -v^2)
    \Xi_a}{(b^2 - a^2)\Delta_v} \nonumber \\
&+& \frac{ 2 L
    (a^2 - v^2) (b^2 - v^2)}{v^2 \Delta_v V(v)} - \frac{2 M (r^2
  + a^2) (r^2 + b^2) }{r^2 \Delta_r U(r)}, \nonumber \\ 
(r^2 + v^2) g^{t \phi_i} &=&  \frac{2 a (b^2 - v^2) L }{ v^2 V(v)} - \frac{2 a
  (r^2 + b^2)  M}{r^2 U(r)}, \nonumber \\
(r^2 + v^2) g^{t \tilde{\phi}_i} &=&  \frac{2 b (a^2 - v^2) L }{v^2 V(v)} -
\frac{2 b (r^2 + a^2)  M}{r^2 U(r)}, \nonumber \\
(r^2 + v^2) g^{\phi_i \tilde{\phi}_j} &=&  \frac{2 a b \Delta_v
  L }{ v^2 V(v)} - \frac{2 a b \Delta_r  M}{r^2 U(r)}, \nonumber \\
(r^2 + v^2) g^{\phi_i \phi_j} &=& \frac{\Xi_a (b^2 - a^2)}{\nu_i^2}
\Bigg(\frac{1}{r^2 + a^2} - \frac{1}{a^2 - v^2} \Bigg) \delta^{i j}  +
\frac{2 a^2  (b^2 - v^2) \Delta_v L }{ v^2 (a^2 - v^2) V(v)}
 \nonumber \\
 &-& \frac{2 a^2 (r^2 + b^2) \Delta_r  M}{r^2 (r^2 + a^2) U(r)},
 \nonumber \\
(r^2 + v^2) \, g^{\tilde{\phi}_i \tilde{\phi}_j} &=& \frac{\Xi_b (a^2 -
  b^2)}{\tilde{\nu}_i^2} \Bigg(\frac{1}{r^2 + b^2} 
- \frac{1}{b^2 - v^2} \Bigg) \delta^{i j}  +  \frac{2 b^2 
  (a^2 - v^2) \Delta_v L }{ v^2 (b^2 - v^2) V(v)} \nonumber \\
  &-& \frac{2 b^2 (r^2 + a^2) \Delta_r M}{r^2 (r^2 + b^2) U(r)},
\nonumber \\
(r^2 + v^2) g^{r r} &=& \frac{U(r)}{(r^2 + a^2)^{p - 1} (r^2 + b^2)^{q
    - 1} }, \nonumber \end{eqnarray}
\begin{eqnarray}
(r^2 + v^2) g^{v v} &=& \frac{V(v)}{(a^2 - v^2)^{p - 1} (b^2 - v^2)^{q
    - 1}}, \nonumber \\
(r^2 + v^2) g^{\alpha_i \alpha_j} &=& \frac{\Xi_a (a^2 - b^2)}{
  \prod_{k = 1}^{i -1} \sin^2 \alpha_k} \Bigg( \frac{1}{a^2 - v^2} -
\frac{1}{r^2  + a^2} \Bigg) \delta^{i j}, \nonumber \\
(r^2 + v^2) g^{\beta_i \beta_j} &=& \frac{\Xi_b (b^2 - a^2)}{
  \prod_{k = 1}^{i - 1} \sin^2 \beta_k} \Bigg( \frac{1}{b^2 - v^2} -
\frac{1}{r^2  + b^2} \Bigg) \delta^{i j}.
\end{eqnarray}
Comparing the five-dimensional inverse to that given in \cite{dkl}
(having set the charge, $q$ to zero), we see that there is complete
agreement when $L=0$ \footnote{In five dimensions, the NUT parameter
  is trivial anyway.} and the slightly different definitions have been
taken into account. The dependence of $g^{\phi_i \phi_j}$ on $\nu_i^2$
may look troublesome, but in fact it does not affect the separation of
the Hamilton-Jacobi equation or the Klein-Gordon equation.

\vskip 0.2cm
\noindent
We have introduced some new notation. The latitudes $\nu_i$ and
$\tilde{\nu}_j$ have been written in terms of $\alpha_i$ and $\beta_j$
according to the following equations \footnote{Since there are only
  $p-1$ independent values of $\nu_i$ we set $\alpha_p =
  0$. Similarly, we set $\beta_q = 0$.},
\begin{equation} \label{alpha:beta}
\nu_i = \cos \alpha_{p - i + 1} \prod_{k = 1}^{p -i} \sin \alpha_k,
\quad \textrm{and} \quad \tilde{\nu}_j = \cos \beta_{q - j + 1}
\prod_{m = 1}^{q - j} \sin \beta_m.
\end{equation}


\subsection{The Hamilton-Jacobi equation in odd dimensions}
In this section, we shall prove that the Hamilton-Jacobi equation has a
separable solution. We shall also determine the Killing tensor which
gives rise to the extra constant of the motion required for the
spacetime to be geodesically integrable.   

\vskip 0.2cm
\noindent
The Hamilton-Jacobi equation is
\begin{equation} \label{jacobi}
\frac{ \partial S}{\partial \tau} + \frac{1}{2} g^{\mu \nu} \frac{
  \partial S}{\partial x^{\mu}}\frac{\partial S}{\partial x^{\nu}} = 0.
\end{equation}
We shall assume a separable solution for Hamilton's Principal Function:
\begin{equation} \label{S:ansatz}
S = \frac{1}{2} m^2 \tau - E t + \sum_{i=1}^p \Phi_i \phi_i +
\sum_{j=1}^q \tilde{\Phi}_j \tilde{\phi}_j + S_r(r) + S_v(v) +
\sum_{i=1}^{p - 1} S_{\alpha_i} (\alpha_i) + \sum_{j=1}^{q - 1}
S_{\beta_j} (\beta_j).  
\end{equation}
Inserting this ansatz into the Hamilton-Jacobi equation and
multiplying through by $2(r^2 + v^2)$, we find the following equation:
\begin{eqnarray} \label{hj}
-m^2 (r^2 + v^2) &=& \Bigg( \frac{a^2 b^2 g^2 + r^2 ( g^2 (a^2 + b^2)
    -1)}{\Delta_r } + \frac{b^2 (a^2 - v^2) \Xi_b - a^2 (b^2 -v^2)
    \Xi_a}{(b^2 - a^2)\Delta_v} \nonumber \\
&+& \frac{ 2 L
    (a^2 - v^2) (b^2 - v^2)}{v^2 \Delta_v V(v)} - \frac{2 M (r^2
  + a^2) (r^2 + b^2) }{r^2 \Delta_r U(r)} \Bigg) E^2 \nonumber \\
&-& 4 a E \Bigg( \frac{(b^2 - v^2) L }{ v^2 V(v)} - \frac{
  (r^2 + b^2)  M}{r^2 U(r)} \Bigg) \sum_{i=1}^p \Phi_i \nonumber \end{eqnarray}
\begin{eqnarray}
&-& 4 b E \Bigg(
  \frac{ (a^2 - v^2) L }{ v^2 V(v)} - \frac{
  (r^2 + a^2)  M}{r^2 U(r)} \Bigg) \sum_{j=1}^q \tilde{\Phi}_j \nonumber \\
&+& 4 a b \Bigg(\frac{ \Delta_v L }{ v^2 V(v)} - \frac{ \Delta_r
    M}{r^2 U(r)} \Bigg) \sum_{i=1}^p \sum_{j=1}^q
  \Phi_i \tilde{\Phi}_j\nonumber \\
&+& \Xi_a (b^2 - a^2) \Bigg(\frac{1}{r^2 + a^2} - \frac{1}{a^2 - v^2}
  \Bigg) \sum_{i=1}^p  \Bigg( \frac{\Phi_i^2}{\nu_i^2}  + \frac{1}{
  \prod_{k = 1}^{i -1} \sin^2 \alpha_k}  \Bigg( \frac{ d
  S_{\alpha_i}}{d \alpha_i} \Bigg)^2 \Bigg) \nonumber \\
&+& \Xi_b (a^2 - b^2) \Bigg(\frac{1}{r^2 + b^2} - \frac{1}{b^2 - v^2}
  \Bigg) \sum_{j=1}^q  \Bigg(\frac{\tilde{\Phi}_j^2}{\tilde{\nu}_j^2}
  + \frac{1}{\prod_{k = 1}^{j -1} \sin^2 \beta_k}  \Bigg( \frac{ d
  S_{\beta_j}}{d \beta_j} \Bigg)^2 \Bigg)
\nonumber 
\end{eqnarray}
\begin{eqnarray}
&+& 2 a^2 \Bigg(  \frac{(b^2 - v^2) \Delta_v L }{ v^2 (a^2 - v^2)
    V(v)} -  \frac{(r^2 + b^2) \Delta_r M}{r^2 (r^2 + a^2) U(r)}
    \Bigg) \sum_{i=1}^p \sum_{j=1}^p \Phi_i \Phi_j\nonumber \\  
&+& 2 b^2 \Bigg(  \frac{(a^2 - v^2) \Delta_v L }{ v^2 (b^2 - v^2)
    V(v)} -  \frac{(r^2 + a^2) \Delta_r M}{r^2 (r^2 + b^2) U(r)}
    \Bigg) \sum_{i=1}^q \sum_{j=1}^q \tilde{\Phi}_i \tilde{\Phi}_j
    \nonumber \\  
&+& \frac{U(r)}{ (r^2 + a^2)^{p-1} (r^2 + b^2)^{q-1}} \Bigg( \frac{d
  S_r}{d r} \Bigg)^2  \nonumber \\
&+& \frac{V(v)}{ (a^2 - v^2)^{p-1} (b^2 - v^2)^{q-1}} \Bigg( \frac{d
  S_v}{d v} \Bigg)^2. 
\end{eqnarray}
The terms involving the $\alpha_i$ and $\beta_j$ can be separated off
and set equal to a constant, that is \footnote{Note that $S_{\alpha_p}
  = S_{\beta_q} \equiv 0$.}
\begin{eqnarray} \label{alphaterms}
\sum_{i=1}^p \Bigg(\frac{\Phi_i^2}{\nu_i^2}  + \frac{1}{
  \prod_{k = 1}^{i -1} \sin^2 \alpha_k}  \Bigg( \frac{ d
  S_{\alpha_i}}{d \alpha_i} \Bigg)^2 \Bigg) &=& J_1^2, \nonumber \\
\textrm{and} \quad \sum_{j=1}^q \Bigg(
  \frac{\tilde{\Phi}_j^2}{\tilde{\nu}_j^2} + \frac{1}{\prod_{k = 1}^{j
  -1} \sin^2 \beta_k}  \Bigg( \frac{ d S_{\beta_j}}{d \beta_j}
  \Bigg)^2 \Bigg)&=& L_1^2. 
\end{eqnarray}
We shall deal with these terms in more detail later. Returning to the
Hamilton-Jacobi equation (\ref{hj}), we can now separate the $r$ and
$v$ dependence, obtaining the two equations:
\begin{eqnarray} \label{sep:r} 
-K  &=& m^2 r^2 + \Bigg(\frac{a^2 b^2 g^2 + r^2 ( g^2 (a^2 + b^2)
    -1)}{\Delta_r } - \frac{2 M (r^2
 + a^2) (r^2 + b^2) }{r^2 \Delta_r U(r)} \Bigg) E^2 \nonumber \\
&+&  \frac{4 a (r^2 + b^2)  M}{r^2 U(r)} \sum_{i=1}^p E \Phi_i
 + \frac{ 4 b (r^2 + a^2)  M}{r^2 U(r)} \sum_{j=1}^q E
  \tilde{\Phi}_j\nonumber \\ 
&-&  \frac{ 4 a b \Delta_r M}{r^2 U(r)} \sum_{i=1}^p \sum_{j=1}^q
  \Phi_i \tilde{\Phi}_j + \frac{ \Xi_a (b^2 - a^2) }{r^2 + a^2} J_1^2
  + \frac{ \Xi_b (a^2 - b^2)}{r^2 + b^2} L_1^2 \nonumber  \end{eqnarray}
\begin{eqnarray}
&-& \frac{2 a^2 (r^2 + b^2) \Delta_r M}{r^2 (r^2 + a^2) U(r)}
 \sum_{i=1}^p \sum_{j=1}^p  \Phi_i \Phi_j - \frac{2 b^2 (r^2 + a^2)
 \Delta_r M}{r^2 (r^2 + b^2) U(r)}  \sum_{i=1}^q \sum_{j=1}^q
  \tilde{\Phi}_i \tilde{\Phi}_j \nonumber \\ 
&+& \frac{U(r)}{ (r^2 + a^2)^{p-1} (r^2 + b^2)^{q-1}} \Bigg( \frac{d
  S_r}{d r} \Bigg)^2,
\end{eqnarray}
and
\begin{eqnarray} \label{sep:v}
K &=& m^2 v^2 + \Bigg( \frac{b^2 (a^2 - v^2) \Xi_b - a^2 (b^2 -v^2)
    \Xi_a}{(b^2 - a^2)\Delta_v} +
\frac{2 L (a^2 - v^2) (b^2 - v^2)}{v^2 \Delta_v V(v)}
\Bigg) E^2 \nonumber \\
&-&  \frac{4 a (b^2 - v^2) L }{ v^2 V(v)} \sum_{i=1}^p E \Phi_i -
\frac{4 b (a^2 - v^2) L }{ v^2 V(v)} \sum_{j=1}^q E \tilde{\Phi}_j
\nonumber \\ 
&+& \frac{4 a b \Delta_v L }{ v^2 V(v)}  \sum_{i=1}^p \sum_{j=1}^q
  \Phi_i \tilde{\Phi}_j - \frac{\Xi_a (b^2 - a^2)}{a^2 - v^2} J_1^2 -
  \frac{\Xi_b (a^2 - b^2)}{b^2 - v^2} L_1^2 \nonumber  
\end{eqnarray}
\begin{eqnarray}
&+&   \frac{2 a^2 (b^2 - v^2) \Delta_v L }{ v^2 (a^2 - v^2) V(v)}
  \sum_{i=1}^p \sum_{j=1}^p \Phi_i \Phi_j +   \frac{2 b^2 (a^2 - v^2)
    \Delta_v L }{ v^2 (b^2 - v^2) V(v)}  \sum_{i=1}^q \sum_{j=1}^q
  \tilde{\Phi}_i \tilde{\Phi}_j \nonumber \\ 
&+& \frac{V(v)}{ (a^2 - v^2)^{p-1} (b^2 - v^2)^{q-1}} \Bigg( \frac{d
  S_v}{d v} \Bigg)^2. 
\end{eqnarray} 
The last thing to show is that the $\alpha_i$ and $\beta_j$ terms can
also be separated. One can easily show that equation
(\ref{alphaterms}) can be separated, giving rise to the following
equations for the $S_{\alpha_i}$ \cite{vs}:
\begin{equation} \label{alphasep1}
\Bigg( \frac{d S_{\alpha_{p-1}}}{ d \alpha_{p-1}} \Bigg)^2 = J_{p
  -1}^2 - \frac{\Phi_1^2}{\sin^2 \alpha_{p - 1}} -
\frac{\Phi_2^2}{\cos^2 \alpha_{p - 1}},
\end{equation}
and 
\begin{equation} \label{alphasep2}
\Bigg( \frac{d S_{\alpha_k}}{ d \alpha_k} \Bigg)^2 = J_k^2 -
\frac{J_{k+1} ^2}{\sin^2 \alpha_k} -
\frac{\Phi_{p - k +1}^2}{\cos^2 \alpha_k},
\end{equation}
for $k = 1,2, \ldots, p-2$. Similarly, for the $\beta_i$ terms, we
find:
\begin{equation}
\Bigg( \frac{d S_{\beta_{q-1}}}{ d \beta_{q-1}} \Bigg)^2 = L_{q
  -1}^2 - \frac{\tilde{\Phi}_1^2}{\sin^2 \beta_{q - 1}} -
\frac{\tilde{\Phi}_2^2}{\cos^2 \beta_{q - 1}},
\end{equation}
and 
\begin{equation}
\Bigg( \frac{d S_{\beta_k}}{ d \beta_k} \Bigg)^2 = L_k^2 -
\frac{L_{k+1} ^2}{\sin^2 \beta_k} -
\frac{\tilde{\Phi}_{q - k +1}^2}{\cos^2 \beta_k},
\end{equation}
for $k = 1,2, \ldots, q-2$. This confirms that the separable ansatz
(\ref{S:ansatz}) is true and that the Hamilton-Jacobi equation is
indeed separable. Thus, the Kerr-AdS-NUT spacetime is geodesically
integrable in odd dimensions. 

\vskip 0.2cm
\noindent
The irreducible Killing tensor can be extracted from the $v$-dependent
separation equation, (\ref{sep:v}). We shall define the Killing tensor
by the equation $K = K^{\mu \nu} p_{\mu} p_{\nu}$, where $p_{\mu}$ is
the canonical momentum. Recalling that $m^2 = - g^{\mu \nu} p_{\mu}
p_{\nu}$, we find
\begin{eqnarray} \label{killingtensor}
K_o^{\mu \nu} &=& - v^2 g^{\mu \nu} + \Bigg( \frac{b^2 (a^2 - v^2) \Xi_b
    - a^2 (b^2 -v^2) \Xi_a}{(b^2 - a^2)\Delta_v} + \frac{2 L (a^2 -
  v^2) (b^2 - v^2)}{v^2 \Delta_v V(v)} \Bigg) \delta_t^{\mu}
\delta_t^{\nu} \nonumber \\ 
&-& \frac{2 a (b^2 - v^2) L}{v^2 V(v)} \sum_{i=1}^p \Big( \delta_t^{\mu}
  \delta_{\phi_i}^{\nu} + \delta_{\phi_i}^{\mu} \delta_t^{\nu} \Big) -
  \frac{2 b (a^2 - v^2) L}{v^2 V(v)} \sum_{j=1}^q \Big( \delta_t^{\mu} 
  \delta_{\tilde{\phi}_j}^{\nu} + \delta_{\tilde{\phi}_j}^{\mu}
  \delta_t^{\nu} \Big) \nonumber \\ 
&+& \frac{2 a b \Delta_v L}{v^2 V(v)} \sum_{i=1}^p \sum_{j=1}^q \Big(
\delta_{\phi_i}^{\mu} \delta_{\tilde{\phi}_j}^{\nu} +
\delta_{\tilde{\phi}_j}^{\mu} \delta_{\phi_i}^{\nu} \Big) +
  \frac{V(v)}{(a^2 - v^2)^{p-1} (b^2 - v^2)^{q-1}} \delta_v^{\mu}
  \delta_v^{\nu} \nonumber \\
&+& \frac{a^2 (b^2 - v^2) \Delta_v L }{v^2 (a^2 - v^2) V(v)}
\sum_{i=1}^p \sum_{i=1}^q \Big( \delta_{\phi_i}^{\mu}
\delta_{\phi_j}^{\nu} + \delta_{\phi_j}^{\mu} \delta_{\phi_i}^{\nu}
\Big) -
\frac{\Xi_b (b^2 - a^2)}{a^2 - v^2} J_1^{\mu \nu}\nonumber \\
 &+& \frac{b^2 (a^2 - v^2) \Delta_v L }{v^2 (b^2 - v^2) V(v)}
\sum_{i=1}^q \sum_{j=1}^q \Big( \delta_{\tilde{\phi}_i}^{\mu}
\delta_{\tilde{\phi}_j}^{\nu} + \delta_{\tilde{\phi}_j}^{\mu}
\delta_{\tilde{\phi}_i}^{\nu} \Big) - \frac{\Xi_a (a^2 -
  b^2)}{b^2 - v^2} L_1^{\mu \nu} ,
\end{eqnarray}
where the tensors, $J_1^{\mu \nu}$ and $L_1^{\mu \nu}$ are the
reducible Killing tensors associated to the $\alpha$ and $\beta$ 
terms in the separation \cite{vsp}. Formulae for these are derived in
the appendix (\ref{Jmunu}, \ref{Lmunu}). We have confirmed that this
Killing tensor reduces to that given in equation (11) of \cite{dkl} when the
appropriate limits are taken.  Unlike the Killing tensor given in \cite{vs},
(\ref{killingtensor}) has a smooth limit as $g \rightarrow 0$. This is
merely an artifact of the non-uniqueness of separating the Hamilton-Jacobi
equation, or equivalently, the $r$ and $v$ dependence of the inverse
metric. By adding appropriate outer products of Killing vectors, the
$L=0$ limit of (\ref{killingtensor}) can be made to agree with that
given in \cite{vs}. 

\vskip 0.2cm
\noindent
Since there are only two rotation parameters, there are two sets of
identical planes of rotation. These give rise to extra Killing 
vectors whose action is to shift between the planes of a given rotation
parameter. These Killing vectors, along with the Killing tensor
presented above provide the extra constants of motion needed to ensure
that the geodesic equation is integrable.   

  
\subsection{The Klein-Gordon equation in odd dimensions}

The Klein-Gordon equation governs the quantum field theory of massive,
spinless particles on this background. It can be written as
\begin{equation}
\frac{1}{\sqrt{-\textrm{det} g}} \, \partial_{\mu} ( \sqrt{-\textrm{det}
  g} \, g^{\mu \nu}\, \partial_{\nu} \Phi) = m^2 \Phi.
\end{equation} 
The separability of the Klein-Gordon equation depends, crucially, on
the coordinate dependence of the metric determinant being factorisable. We
find that the metric determinant is given by
\begin{equation}\label{detodd}
\textrm{det} \, g = -\frac{r^2 v^2 (r^2 + a^2)^{p-1} (r^2 + b^2)^{q-1} (a^2 -
  v^2)^{p-1} (b^2 - v^2)^{q-1} \rho^{2n-2}
  \omega^{2n-2}}{\Xi_a^{2p} \Xi_b^{2q} (b^2 - a^2)^{2p+2q-2}} \nonumber
\end{equation} 
\begin{equation}
\times \hat{\gamma}_{p-1} \hat{\gamma}_{q-1} \prod_{i=1}^p \nu_i^2
\prod_{j=1}^q \tilde{\nu}_j^2, 
\end{equation} 
where $\hat{\gamma}_m$ is the determinant of the metric on a unit
$m$-sphere. It is important to note that the $\rho^{2n-2}
\omega^{2n-2}$ term provides a factor of $r^2 + v^2$ in 
$\sqrt{-\textrm{det} g}$, since this allows the inverse metric to separate
additively. The rest of the determinant is factorisable into functions of $r$,
$v$ $\alpha_i$ and $\beta_j$.  Thus, the Klein-Gordon equation in odd
dimensions will have separable solutions of the form: \footnote{The
  dependence on the latitude coordinates $\alpha$ and $\beta$ may
  appear to cause problems because there is $\alpha$ dependence in the
  diagonal part of $g^{\phi_i \phi_j}$. However, the equations
  involving the latitudes are similar to those in the Hamilton-Jacobi
  equation and can be separated in exactly the same way.}  
\begin{equation}
\Phi = e^{-i \omega t} e^{i \sum_{i=1}^p m_i \phi_i} e^{i
  \sum_{j=1}^q n_j \tilde{\phi}_j} R(r) W(v) \prod_{k=1}^{p-1} A_k(\alpha_k)
\prod_{l=1}^{q-1} B_l(\beta_l) .
\end{equation}


\section{Even Dimensional Kerr-AdS-NUT Solutions}
The metric for even dimensions, $D = 2 n$, can be obtained from the
Kerr-AdS metric by splitting the rotation parameters into two
sets, with one set taking the value $a$ and the other set taking the
value $0$ and then including the NUT parameter as before. The
latitude coordinates can be split up in the same way as in
(\ref{lat}), introducing a coordinate, $\theta$. To simplify the
equations, we replace $\theta$ with the coordinate $v$ defined by
\begin{equation}
v^2 = a^2 \cos^2 \theta,
\end{equation}
which then gives the metric
\begin{eqnarray}
d s^2 &=& - \frac{ ( 1+ g^2 r^2)(1 - g^2 v^2)}{\Xi_a} d t^2 +
\frac{\rho^{2 n -3} d r^2}{U} + \frac{\omega^{2 n -3} d v^2}{V}
\nonumber \\
&+& \frac{2 M r}{\rho^{2 n -3}} \Bigg( \frac{ ( 1- g^2 v^2)}{\Xi_a} d t
- \mathcal{A} \Bigg)^2 - \frac{2 L v}{\omega^{2 n -3}} \Bigg( \frac{ (
  1 + g^2 r^2)}{\Xi_a} d t - \tilde{\mathcal{A}} \Bigg)^2 \nonumber \\
&+& \frac{ (r^2 + a^2)(a^2 - v^2)}{a^2 \Xi_a} \sum_{i=1}^p ( d \nu_i^2 +
\nu_i^2 d \phi_i^2) + \frac{r^2 v^2}{a^2} \Bigg( d \tilde{\nu}_{q+1}^2
+ \sum_{j=1}^q( d \tilde{\nu}_j^2 + \tilde{\nu}_j^2 d
\tilde{\phi}_j^2) \Bigg),
\end{eqnarray}
where
\begin{eqnarray}
\mathcal{A} &=& \frac{ a^2 - v^2}{a \Xi_a} \sum_{i=1}^p \nu_i^2 d
\phi_i, \quad \quad \tilde{\mathcal{A}} = \frac{ r^2 + a^2}{a \Xi_a}
\sum_{i=1}^p \nu_i^2 d \phi_i, \nonumber \\
U &=& (1+g^2 r^2)(r^2 + a^2)^p r^{2q} - 2 M r,\nonumber \\
V &=& (1- g^2 v^2)(a^2 - v^2)^p v^{2q} - 2 L v,\nonumber \\
\rho^{2 n -3} &=& (r^2 + v^2)(r^2 + a^2)^{p - 1} r^{2q},\nonumber \\
\omega^{2 n -3} &=& (r^2 + v^2)(a^2 - v^2)^{p - 1} v^{2q},
\end{eqnarray}
and $\Xi_a$ is given by the same equation as in the odd dimensional
case. The relationship between $p$, $q$ and $n$ is now $p + q + 1 =
n$. The definitions of $\Delta_r$ and $\Delta_v$ are the same as in
the odd dimensional case.  

\vskip 0.2cm
\noindent
Again, in order to show that the Hamilton-Jacobi equation is
separable, and hence that the geodesic equation is integrable, we must
show that the inverse metric can be split up in a particular
way. Computing the inverse metric, we find
\begin{eqnarray}
(r^2 + v^2) g^{t t} &=& -\frac{v^2 \Xi_a}{\Delta_v} - \frac{r^2
    \Xi_a}{\Delta_r} + \frac{2 L v (a^2 - v^2)}{\Delta_v V(v)} -
    \frac{2 M r (r^2 + a^2)}{\Delta_r U(r)}, \nonumber \\
(r^2 + v^2) g^{t \phi_i} &=& \frac{2 a L v}{V(v)} - \frac{2 a M
    r}{U(r)},  \nonumber \\
(r^2 + v^2) g^{\phi_i \phi_j} &=& - \frac{a^2
    \Xi_a}{\nu_i^2} \Bigg( \frac{1}{r^2 + a^2} - \frac{1}{a^2 - v^2}
    \Bigg) \delta^{i j}  + \frac{2 a^2 L v \Delta_v}{(a^2 - v^2) V(v)}
  - \frac{2 a^2 M r \Delta_r}{(r^2  + a^2) U(r)}  \nonumber \\
(r^2 + v^2) g^{\tilde{\phi}_i \tilde{\phi}_j} &=&
  \frac{a^2}{\tilde{\nu}_j^2}  \Bigg( \frac{1}{r^2} + \frac{1}{v^2}
  \Bigg) \delta^{i j}, \nonumber \\ 
(r^2 + v^2) g^{r r} &=& \frac{U(r)}{(r^2 + a^2)^{p - 1} r^{2q}},
  \nonumber \\
(r^2 + v^2) g^{v v} &=& \frac{V(v)}{(a^2 - v^2)^{p - 1} v^{2q}},
  \nonumber \\
(r^2 + v^2) g^{\alpha_i \alpha_j} &=& \frac{a^2 \Xi_a}{\prod_{k=1}^{i-1}
    \sin^2 \alpha_k} \Bigg( \frac{1}{a^2 -
    v^2} - \frac{1}{r^2 + a^2} \Bigg) \delta^{i j}, \nonumber \\
(r^2 + v^2) g^{\beta_i \beta_j} &=& \frac{a^2}{\prod_{k=1}^{i-1}
    \sin^2 \beta_k} \Bigg( \frac{1}{r^2} + \frac{1}{v^2} \Bigg)
    \delta^{i j},
\end{eqnarray}
where, as before, we have introduced latitude coordinates, $
\{\alpha_i \}$ and $\{ \beta_j \}$, with $ \alpha_i$ defined
as in (\ref{alpha:beta}), and $\beta_j$ defined by the slightly
modified equation  
\begin{equation}
\tilde{\nu}_j = \cos \beta_{q - j + 2} \prod_{m = 1}^{q + 1 - j} \sin
\beta_m.
\end{equation}
As in the odd dimensional case, we have set $\alpha_p = \beta_{q+1} =
0$ to impose the constraints on the $\{ \nu_i \}$ and the $\{
\tilde{\nu}_j \}$. 

\vskip 0.2cm
\noindent
This inverse metric is simpler than the odd-dimensional case because
there is no coupling between $t$ and the $\{ \tilde{\phi}_i \}$. Again, the
$r$ and $v$ dependence separates and this will allow the separation
of the Hamilton-Jacobi equation.


\subsection{The Hamilton-Jacobi equation in even dimensions}
We consider an ansatz for Hamilton's Principal Function, $S$, that is
separable:
\begin{equation}
S = \frac{1}{2} m^2 \tau - E t + \sum_{i=1}^p \Phi_i \phi_i +
\sum_{j=1}^{q} \tilde{\Phi}_j \tilde{\phi}_j + S_r(r) + S_v(v) +
\sum_{i=1}^{p-1} S_{\alpha_i}(\alpha_i) + \sum_{j=1}^q S_{\beta_j}
(\beta_j). 
\end{equation}
Substituting this ansatz into the Hamilton-Jacobi equation,
(\ref{jacobi}), and multiplying through by $r^2 + v^2$ we obtain
\begin{eqnarray} \label{hjeven}
-m^2 (r^2 + v^2) &=& \Bigg( \frac{- r^2 \Xi_a}{\Delta_r } + \frac{-
    v^2 \Xi_a}{\Delta_v} +  \frac{ 2 L v
    (a^2 - v^2) }{\Delta_v V(v)} - \frac{2 M r (r^2
  + a^2) }{ \Delta_r U(r)} \Bigg) E^2 \nonumber 
\end{eqnarray}
\begin{eqnarray}
&-& 4 a E \Bigg( \frac{L v}{V(v)} - \frac{M r}{U(r)} \Bigg)
    \sum_{i=1}^p \Phi_i \nonumber \\
&-& \Xi_a  a^2 \Bigg(\frac{1}{r^2 + a^2} - \frac{1}{a^2 - v^2}
  \Bigg) \sum_{i=1}^p  \Bigg( \frac{\Phi_i^2}{\nu_i^2}  + \frac{1}{
  \prod_{k = 1}^{i -1} \sin^2 \alpha_k}  \Bigg( \frac{ d
  S_{\alpha_i}}{d \alpha_i} \Bigg)^2 \Bigg) \nonumber \\
&+& a^2 \Bigg(\frac{1}{r^2} + \frac{1}{v^2}
  \Bigg) \sum_{j=1}^q  \Bigg(\frac{\tilde{\Phi}_j^2}{\tilde{\nu}_j^2}
  + \frac{1}{\prod_{k = 1}^{j -1} \sin^2 \beta_k}  \Bigg( \frac{ d
  S_{\beta_j}}{d \beta_j} \Bigg)^2 \Bigg)
\nonumber \\ 
&+& 2 a^2 \Bigg(  \frac{L v \Delta_v}{(a^2 - v^2) V(v)} -  \frac{M r
    \Delta_r}{ (r^2 + a^2) U(r)}
    \Bigg) \sum_{i=1}^p \sum_{j=1}^p \Phi_i \Phi_j\nonumber \\  
&+& \frac{U(r)}{ (r^2 + a^2)^{p-1} r^{2q}} \Bigg( \frac{d
  S_r}{d r} \Bigg)^2  \nonumber \\
&+& \frac{V(v)}{ (a^2 - v^2)^{p-1} v^{2q}} \Bigg( \frac{d
  S_v}{d v} \Bigg)^2. 
\end{eqnarray}
As in the odd dimensional case, one can separate out the $\alpha$
dependent terms, obtaining (\ref{alphasep1}) and
(\ref{alphasep2}) so we can replace the $\alpha$ dependent terms in
(\ref{hjeven}) with a constant, $J_1^2$.    

\vskip 0.2cm
\noindent
The $\beta$ dependent terms separate in a similar way to odd
dimensional case, but there is a slight difference due to the
number of latitudes being the same as the number of azimuthal angles,
instead of one fewer. Setting
\begin{eqnarray}
\Bigg( \frac{ d S_{\beta_q}}{d \beta_q} \Bigg)^2 &=& L_q^2 -
\frac{\tilde{\Phi}^2_1}{\sin^2 \beta_q} -
\frac{\tilde{\Phi}^2_2}{\cos^2 \beta_q}, \nonumber \\
\Bigg( \frac{ d S_{\beta_1}}{d \beta_1} \Bigg)^2 &=& L_1^2 -
\frac{L^2_2}{\sin^2 \beta_1} ,  
\end{eqnarray}
and 
\begin{equation}
\Bigg( \frac{ d S_{\beta_k}}{d \beta_k} \Bigg)^2 = L_k^2 -
\frac{L_{k+1}^2}{\sin^2 \beta_k} -
\frac{\tilde{\Phi}^2_{q+2-k}}{\cos^2 \beta_k}, 
\end{equation}
for $k=2, \ldots, q-1$, allows the $\beta$ dependent terms in
(\ref{hjeven}) to be replaced by a constant, $L_1^2$.    

\noindent
We can now separate (\ref{hjeven}) into terms depending only on $r$ and
$v$.
\begin{eqnarray} 
-K &=& m^2 r^2 - \Bigg( \frac{r^2 \Xi_a}{\Delta_r } + \frac{2 M r (r^2
  + a^2) }{ \Delta_r U(r)} \Bigg) E^2 + \frac{4 a M r}{U(r)}
    \sum_{i=1}^p E \Phi_i \nonumber \end{eqnarray}
\begin{eqnarray}
&-& \frac{\Xi_a  a^2}{r^2 + a^2} 
   J_1^2 +  \frac{a^2}{r^2} L_1^2 - \frac{2 a^2 M r \Delta_r}{ (r^2 +
  a^2) U(r)} \sum_{i=1}^p \sum_{j=1}^p \Phi_i \Phi_j\nonumber \\  
&+& \frac{U(r)}{ (r^2 + a^2)^{p-1} r^{2q}} \Bigg( \frac{d
  S_r}{d r} \Bigg)^2  , 
\end{eqnarray}
and
\begin{eqnarray} \label{vdep:even}
K &=& m^2 v^2 - \Bigg(  \frac{v^2 \Xi_a}{\Delta_v} -  \frac{ 2 L v
    (a^2 - v^2) }{\Delta_v V(v)} \Bigg) E^2 - \frac{4 a L v}{V(v)}
    \sum_{i=1}^p E \Phi_i \nonumber \\ 
&+& \frac{\Xi_a  a^2}{a^2 - v^2} J_1^2 + \frac{a^2}{v^2} L_1^2
+   \frac{2 a^2 L v \Delta_v}{(a^2 - v^2) V(v)} \sum_{i=1}^p
    \sum_{j=1}^p \Phi_i \Phi_j \nonumber \\
&+& \frac{V(v)}{ (a^2 - v^2)^{p-1} v^{2q}} \Bigg( \frac{d
    S_v}{d v} \Bigg)^2. 
\end{eqnarray}
We have therefore shown that
the Hamilton-Jacobi equation for the AdS-Kerr-NUT spacetime in even
dimensions is separable and that geodesic motion on it is
integrable.   

\vskip 0.2cm
\noindent
We can use (\ref{vdep:even}) to write down an irreducible Killing
tensor for the AdS-Kerr-NUT spacetime in even dimensions:
\begin{eqnarray} \label{kteven}
K_e^{\mu \nu} &=& - v^2 g^{\mu \nu} - \Bigg(  \frac{v^2 \Xi_a}{\Delta_v}
    -  \frac{ 2 L v (a^2 - v^2) }{\Delta_v V(v)} \Bigg) \delta^{\mu}_t
    \delta_t^{\nu} - \frac{2 a L v}{V(v)}
    \sum_{i=1}^p ( \delta^{\mu}_t \delta^{\nu}_{\phi_i} +
    \delta^{\mu}_{\phi_i} \delta_t^{\nu}) \nonumber \\ 
&+& \frac{\Xi_a  a^2}{a^2 - v^2} J_1^{\mu \nu} + \frac{a^2}{v^2}
    L_1^{\mu \nu} +   \frac{a^2 L v \Delta_v}{(a^2 - v^2) V(v)} \sum_{i=1}^p
    \sum_{j=1}^p (\delta^{\mu}_{\phi_i} \delta^{\nu}_{\phi_j} +
    \delta^{\mu}_{\phi_j} \delta^{\nu}_{\phi_i} ) \nonumber \\
&+& \frac{V(v)}{ (a^2 - v^2)^{p-1} v^{2q}} \delta_v^{\mu}
    \delta_v^{\nu} . 
\end{eqnarray}
Again, we have computed the inverse metric in such a way that the
Killing tensor has a smooth limit as $g \rightarrow 0$. The tensors
$J_1^{\mu \nu}$ and  $L_1^{\mu \nu}$ are the reducible Killing tensors arising
from the $\alpha$ and $\beta$ separation and are given in the
appendix, by equations (\ref{Jmunu}) and (\ref{Lmunu}). They take the
same form as in odd dimensions.  


\subsection{The Klein-Gordon equation in even dimensions}
As in the odd-dimensional case, the separability of the Klein-Gordon
equation in even dimensions will depend upon the factorisation of the
determinant of the metric. We find the following result:
\begin{equation} \label{deteven}
\textrm{det} \, g = -\frac{r^{2q} v^{2q} (r^2 + a^2)^{p-1}  (a^2 -
  v^2)^{p-1} \rho^{2n-3} \omega^{2n-3}}{a^{4p+4q-2} \Xi_a^{2p}}
  \hat{\gamma}_{p-1} \hat{\gamma}_q \prod_{i=1}^p \nu_i^2
  \prod_{j=1}^q \tilde{\nu}_j^2.
\end{equation}
Again, this determinant possesses a factor of $(r^2 + v^2)^2$ so we
can exploit the additive separability of the inverse metric. The
remainder then factorises into parts depending on $r$, $v$, $\alpha_i$
and $\beta_j$. As in the odd dimensional case, the potentially
problematic $\alpha_i$ dependence of $g^{\phi_i \phi_j}$ can be
handled in the same manner as when the Hamilton-Jacobi equation was
separated. Therefore, the Klein-Gordon equation in even dimensions
possesses separable solutions of the form
 \begin{equation}
\Phi = e^{-i \omega t} e^{i \sum_{i=1}^p m_i \phi_i} e^{i
  \sum_{j=1}^q n_j \tilde{\phi}_j} R(r) W(v) \prod_{k=1}^{p-1} A_k(\alpha_k)
\prod_{l=1}^{q} B_l(\beta_l).
\end{equation}   


\section{Summary of results}
In this paper, we have studied the recently discovered,
cohomogeneity-2, rotating, AdS black holes with a NUT-like charge. By
computing the inverse metric in all dimensions, and showing that $(r^2
+ v^2) g^{\mu \nu}$ is additively separable as a function of $r$ and
$v$, we were able to demonstrate complete separation of the Hamilton-Jacobi
equation. This non-trivial result means that the geodesic equation is
integrable. The integrability of the geodesic equation can be explained
by a hidden symmetry that gives rise to an extra constant of the
motion which is quadratic in the canonical momenta. This is intimately
tied to the existence of a second rank Killing tensor which we have
found for all dimensions and displayed in equations
(\ref{killingtensor}) and (\ref{kteven}).  

\vskip 0.2cm
\noindent
We also considered the Klein-Gordon equation for this new class of
black holes. By computing the determinant of the metric in both odd
(\ref{detodd}) and even (\ref{deteven}) dimensions, we were able to
show that the combination $\sqrt{- \textrm{det} g} \, g^{\mu \nu}$
factorises and therefore, the Klein-Gordon equation will possess
separable solutions, allowing the quantum field theory of massive,
spinless particles to be readily understood.       


\section{Acknowledgements}
The author wishes to thank Hari K. Kunduri and James Lucietti for many helpful
discussions and comments on this manuscript, and Gary Gibbons for
advice during the later stages of this work. 


\appendix
\section{Killing tensors for the $\alpha$ and $\beta$ sectors} 
The $\alpha_i$ and $\beta_j$ dependent terms in (\ref{hj}) can be
shown to arise as the result of the existence of reducible Killing
tensors in the following way:
\begin{eqnarray} \label{Jmunu}
\sum_{i=1}^p \Bigg( \frac{ \Phi_i^2}{\nu_i^2}  + \frac{ 1 }{\prod_{k
    =1}^{i -1} \sin^2 \alpha_k} \Bigg( \frac{d S_{\alpha_i}}{d
    \alpha_i} \Bigg)^2 \Bigg) &=& \sum_{i,j = 1}^p \Bigg( \frac{
    \delta^{i j} \delta_{\phi_i}^{\mu} 
    \delta_{\phi_j}^{\nu} }{\nu_i^2} + \frac{ \delta^{i j}
    \delta_{\alpha_i}^{\mu} \delta_{\alpha_j}^{\nu} }{\prod_{k
    =1}^{i -1} \sin^2 \alpha_k} \Bigg) \frac{ \partial S}{\partial
    x^{\mu}} \frac{ \partial S}{\partial x^{\nu}} \nonumber \\
&\equiv& J_1^{\mu \nu} \frac{ \partial S}{\partial
    x^{\mu}} \frac{ \partial S}{\partial x^{\nu}},
\end{eqnarray}
with the corresponding $\beta$ result:
\begin{eqnarray} \label{Lmunu}
\sum_{j=1}^q \Bigg( \frac{ \tilde{\Phi}_j^2}{\tilde{\nu}_j^2}  +
    \frac{ 1 }{\prod_{k=1}^{j-1} \sin^2 \beta_k} \Bigg( \frac{d S_{\beta_j}}{d
    \beta_j} \Bigg)^2 \Bigg) &=& \sum_{i,j = 1}^q \Bigg( \frac{
    \delta^{i j} \delta_{\tilde{\phi}_i}^{\mu} 
    \delta_{\tilde{\phi}_j}^{\nu} }{\tilde{\nu}_i^2} + \frac{ \delta^{i j}
    \delta_{\beta_i}^{\mu} \delta_{\beta_j}^{\nu} }{\prod_{k
    =1}^{i -1} \sin^2 \beta_k} \Bigg) \frac{ \partial S}{\partial
    x^{\mu}} \frac{ \partial S}{\partial x^{\nu}} \nonumber \\
&\equiv& L_1^{\mu \nu} \frac{ \partial S}{\partial
    x^{\mu}} \frac{ \partial S}{\partial x^{\nu}}.
\end{eqnarray}


\end{document}